# Enhancing the Electrooptic Effect Using Modulation Instability

**Peter T. S. DeVore, David Borlaug, and Bahram Jalali**[*]

*Department of Electrical Engineering, University of California, Los Angeles, California, USA, 90095*
*Corresponding author: jalali@ucla.edu*

As electronic operating frequencies increase toward the terahertz regime, new electrooptic modulators capable of low-voltage high-frequency operation must be developed to provide the necessary optical interconnects. This letter presents a new concept that exploits modulation instability to enhance the intrinsically weak electrooptic effect, $\chi^{(2)}$. Simulations demonstrate more than 50 times enhancement of electrooptic effect at millimeter wave frequencies leading to a substantial reduction in the required modulation voltage.

Modulation instability (MI) is a nonlinear physical process whereby tiny disturbances spontaneously grow on an otherwise quiet background [1]. First discovered in hydrodynamics where it is known as the Benjamin-Feir instability [2], MI arises in diverse contexts such as sand-dune formation [3], free electron lasers [4], optics [5], and matter-waves [6], and has been described as "one of the most ubiquitous types of instabilities in nature" [7]. This process has been proposed as the mechanism to give rise to oceanic rogue or freak waves [8], giant "walls of water" [9] that appear suddenly even out of calm seas [10]. Discovery of their optical counterpart, optical rogue waves [11], confirmed that specific noise stimulates their generation via modulation instability. Purposefully stimulating the creation of optical rogue waves has led to enhanced bandwidth, stability and brightness [12-15] of broadband coherent light known as supercontinuum [16]. Here, we show how this phenomenon can be exploited to boost the weak electrooptic effect, one of the most fundamental and pressing predicaments in optical communication. In our approach, modulation sidebands stimulate MI in a 3rd order nonlinear optical material placed after the electrooptic device. This effect enhances $\chi^{(2)}$ of the electrooptic material and also compensates for the device related high frequency roll-off. Boosting $\chi^{(2)}$ in one material with the $\chi^{(3)}$-induced MI in another is an intriguing concept that enables low voltage electrooptic modulation at ultrahigh frequencies.

Electrooptic (EO) modulation is the critical bridge that connects electronic computing to optical communication [17, 18]. While today's communication, sensing, and computing systems are predominantly electronic, optics has become indispensible for high-speed communication. Unfortunately, the development of EO modulators has not kept up with the tremendous pace of electronics over the past decades. To be sure, access to broadband optical channels is curtailed by the limited bandwidth of EO modulators that is currently in the 50-100 GHz range [18]. These EO modulator speeds are achieved at the cost of increase in modulation voltage brought about by the decline of electrooptic interaction due to the increased loss of metallic electrodes at high frequencies. As the electronic frontier continues to progress toward the terahertz regime, the modulator bottleneck will become more pronounced. Existing approaches for improving EO modulators are primarily focused on overcoming the RC time constant and increasing the interraction length between optical and electrical waves using traveling wave electrodes [17]. While these have achieved improved frequency responses, the progress has come at the cost of increased electrical drive voltage due to the shorter interaction length required for these bandwidths [19]. As a result, devices can be designed for either low voltage or high frequency, but not both. Another approach for improving EO modulation has been utilizing exotic materials with a strong electroabsorption effect such as transparent conductive oxides [20] and graphene [21, 22].

In this letter, a new concept called the electrooptic booster (EOB) is proposed as a path to broadband low-voltage electrooptic modulation. Our method enhances any modulator including those based on electrorefraction as well as electroabsorption effects. The booster is numerically demonstrated to reduce a Mach-Zehnder modulator's half-wave voltage, known as $V_\pi$, by 17 dB, as well as nearly doubling its bandwidth by exploiting the intrinsince frequency dependence of the MI gain spectrum.

Modulation instability (MI) in optics results from the interplay between Kerr nonlinearity and anomalous group velocity dispersion. The effect can be obtained using linear stability analysis performed on analytical solutions to the nonlinear Schrödinger equation:

$$\frac{\partial A(z,t)}{\partial z} = \sum_{k\geq 2} \frac{i^{k+1}}{k!} \beta_k \frac{\partial^k A(z,t)}{\partial t^k} + i\gamma A(z,t)|A(z,t)|^2 + i\gamma f_R A(z,t) \int_{-\infty}^{t} h_R(t-t')|A(z,t')|^2 dt' - \frac{\alpha}{2} A(z,t) \quad (1)$$

where $A(z,t)$ is the field amplitude (in Watts$^{1/2}$), $\beta_k = \partial^k \beta / \partial \omega_0^k$ are the dispersion coefficients, $\gamma = 3\pi\chi^{(3)}/(4\lambda_0 A_{eff})$ is the Kerr nonlinear coefficient, $A_{eff}$ is the effective modal area, $n$ is the refractive index, $\lambda_0$ is the pump wavelength, $\alpha$ is the loss per unit length, and $f_R$ and $h_R(t)$ are the Raman fractional contribution and response function respectively [1]. Over the electrically relevant pump-sideband frequency difference < 1 THz, the Raman effect and dispersion terms $\beta_k$ with $k > 2$ can be ignored. Though the stochastic nature of temporally-confined MI [23], MI with near-zero $\beta_2$ [24], or when the sidebands grow too large [25-27] involve rather complex dynamics, the initial evolution of MI from continuous-wave radiation with sizable anomalous group velocity dispersion $\beta_2$ can be described analytically. The sideband gain per unit length under this condition without loss or pump depletion is [1]

$$g(\omega) = |\beta_2 \omega|[(4\gamma P/|\beta_2|) - \omega^2]^{1/2} \quad (2)$$

where $\omega$ is the pump-sideband frequency separation and $P$ is the pump optical power (cf. Fig. 1a). Sidebands modulated on a strong optical carrier, acting as the pump, are thus amplified, increasing their modulation depth. The response of EO modulators is characterized by a high frequency roll-off resulting from velocity mismatch between microwave and optical waves in the traveling-wave electrodes, and from electrode microwave losses. The losses are primarily due to the skin effect, and radiative and dielectric loss (from waveguide scattering imperfections and molecular resonances, respectively) [28]. Fortuitously, the increase of MI gain with pump-sideband frequency separation compensates the modulator's high frequency roll-off, enhancing both the bandwidth and the EO sensitivity (cf. Fig 1b). Fig. 1c clearly shows the growth of a 100 GHz sideband (simulation parameters described below). The enhanced modulator utilizing MI to boost modulation extinction in a communication link is shown in Fig. 2.

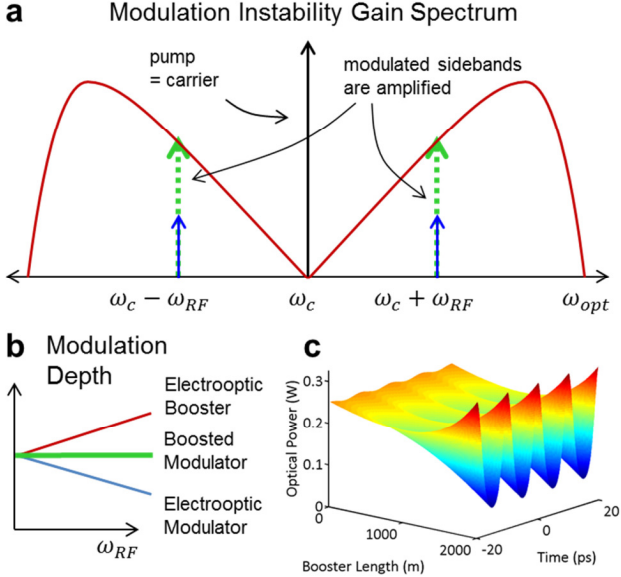

FIG. 1. (color online). (a) Schematic of modulation instability (MI) gain spectrum vs. optical frequency. Sidebands modulated on a strong optical carrier grow as a result of MI. The gain increases with modulation frequency. (b) Schematic of modulation depth vs. RF frequency. Increase in electrooptic booster MI gain with modulation frequency compensates electrooptic modulator roll-off, resulting in a low drive-voltage, wideband modulator. (c) Simulation of MI of a 100 GHz sideband demonstrates the strength of this effect (parameters given in text).

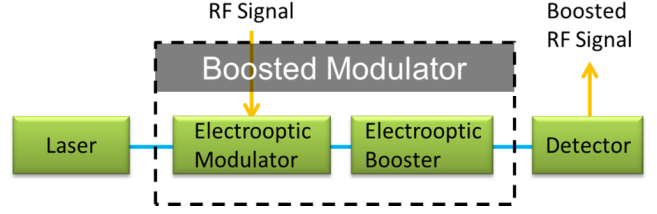

FIG. 2. (color online). Electrooptic booster in a communication link. The conventional link employs only the electrooptic modulator, while the boosted version shown here employs the added electrooptic booster, which uses MI gain to compensate modulation roll-off.

A general approach to characterize EO modulation can be derived from the transfer function for a Mach-Zehnder interferometer with an electrooptic element placed in one arm: $P_{out} = (P/2)\{1 + \cos[\phi_0 + \pi(V/V_\pi)]\}$ [17] where $P$ and $P_{out}$ are the optical powers before and after the modulator, respectively. $V_\pi = sn\lambda_0/(4\chi^{(2)}\Gamma l_{EO})$ [29] is the half-wave voltage (with electrode separation $s$, EO overlap $\Gamma$, and EO interaction length $l_{EO}$), which is the characteristic voltage change required to induce a $\pi$ phase shift between the modulator arms. $\phi_0$ is the static phase offset typically set to quadrature value ($\pi/2$). We consider this modulator fed by a single frequency laser, and driven with a single-tone RF source, $\omega_{RF}$. Finally, the output of the modulator is detected by a square-law photodetector.

The sidebands stimulate MI causing their growth at the expense of the carrier. As the sideband powers increase by $G_{MI}$, the sideband fields increase by $G_{MI}^{1/2}$. As the detected RF current is a result of the carrier-sideband beating, the RF output power is proportional to the square of $G_{MI}^{1/2}$, yielding:

$$P_{RF,out} \propto \frac{1}{V_\pi^2} P_{RF,in} G_{MI} \quad (3)$$

where $V_\pi$ is the inherent Mach-Zehnder half-wave voltage, and $P_{RF,in}$ and $P_{RF,out}$ are the input and output RF powers. From Eqn. (3), the MI-induced growth of $P_{RF,out}$ by $G_{MI}$ is equivalent to $V_\pi$ reduction by $G_{MI}^{1/2}$, yielding the boosted modulator half-wave voltage

$$V_{\pi,eff}(\omega_{RF}) = V_\pi(\omega_{RF}) \frac{L_{Dep}^{1/2}}{G_{MI}^{1/2}(\omega_{RF})} \quad (4)$$

where $L_{Dep} \leq 1$ is the pump (carrier) power depletion factor. Equation (4) demonstrates both key abilities of the EO booster: it effectively reduces the half-wave voltage of modulators, equivalent to increasing $\chi_{eff}^{(2)} =$

$\chi^{(2)}(G_{MI}^{1/2}/L_{Dep}^{1/2})$, as well as flattening the frequency response. The desirable reduction of $V_\pi$ due to pump (carrier) depletion can be understood by recognizing that the transfer of power from the carrier to the sideband increases the ratio of AC to DC signal. For communication applications, equation (4) enhances system performance, as calculated in the boosted shot-noise limited RF signal-to-noise ratio (SNR) [30]:

$$SNR = \frac{R_d Z_m P P_{RF,in}}{4q\Delta f} \frac{\pi^2}{V_{\pi,\text{eff}}^2} \quad (5)$$

where $R_d$ is the detector photodetector responsivity expressed as the ratio of the photogenerated electrical current divided by the incident optical power, $Z_m$ is the electrical impedance of the modulator, $q$ is the elementary charge, and $\Delta f$ is the RF bandwidth.

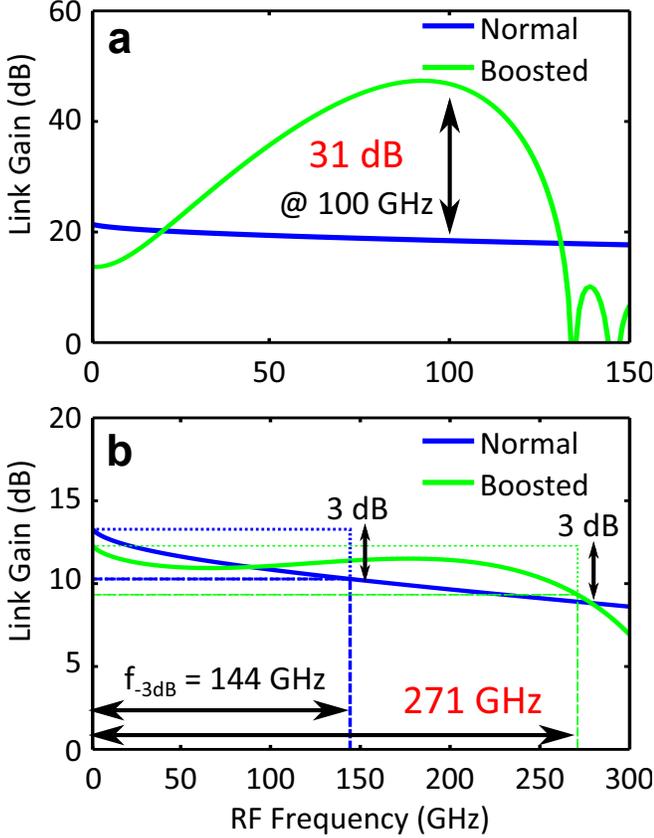

FIG. 3. (color online). Gain $(P_{RF,out}(\omega)/P_{RF,in}(\omega))$ of normal and boosted communication links vs. RF frequency. (a) MI tailored for high gain at 100 GHz (parameters in text). Here the gain is improved 1000 times over the normal link. (b) MI tuned for bandwidth extension ($\beta_2 = -2$ ps$^2$/km, $P = 300$ mW, $l = 500$ m). MI's compensation of the roll-off nearly doubles link bandwidth from 144 GHz to 271 GHz.

The MI effect can be tuned to optimize the Shannon-Hartley channel capacity $C_B = \Delta f \log_2(1 + SNR)$ [31], the absolute upper bound on information transmission bit-rate, with RF bandwidth $\Delta f$. An analytical solution can be found if it is assumed that the noise is independent of MI gain (valid if pump-fluctuation transfer to the sidebands and MI-amplified quantum fluctuations can be neglected [32]), and that $SNR \gg 1$ (satisfied for any useful frequency band), as well as neglecting loss and pump depletion. The channel capacity can then be split into two terms, $C_B = C + \int \log_2(G_{MI}(f)) \, df$, where $C_B$ and $C$ are the boosted and instrinsic channel capacities, respectively. To optimize the performance, one can tune the dispersion yielding the relation for the optimal MI cutoff frequency:

$$\Omega_c = \sqrt{\frac{4\gamma P}{|\beta_2|}} = \omega_{high} \left\{ \frac{4\left[1 - (\omega_{low}/\omega_{high})^6\right]}{3\left[1 - (\omega_{low}/\omega_{high})^2\right]} \right\}^{1/4} \quad (6)$$

where $\omega_{low}$ to $\omega_{high}$ is the electrical RF signal bandwidth. Physically, this choice maximizes the logarithmic power over this band, the quantity of importance for information capacity. It is interesting to note that the form of the pre-boosted $SNR$ has no effect on how the MI gain should be distributed. In the limit $\omega_{low} \to \omega_{high}$ (single frequency), $\Omega_c/\sqrt{2} \to \omega_{high}$, as expected since this places the gain peak at the RF frequency of interest [1]. Substituting the optimal MI dispersion in the channel capacity yields:

$$C_B = C + \left[\left(1 - \frac{\omega_{low}^2}{\Omega_c^2}\right)^{3/2} - \left(1 - \frac{\omega_{high}^2}{\Omega_c^2}\right)^{3/2}\right] \cdot \quad (7)$$
$$2\gamma P l \Omega_c / (3\pi \ln(2))$$

where $l$ is the EO booster length. If we consider the MI values below (which optimizies a frequency band from $f_{low} \approx 0$ to $f_{high} = 191$ GHz), Eqn. (7) predicts a stunning 4.7 bits/s/Hz added capacity in an ideal case. Although optical losses and added noise will reduce this in practice, the analysis does reveal the potential to strongly augment the information capacity of a communication link.

MI is simulated using the setup shown in Fig. 2. A single-frequency 500 mW laser and an ideal RF signal generator with impedance $Z_s = 50$ $\Omega$ feeds a high-speed Mach-Zehnder modulator. The modulator is modeled [17, 28] after Ref. [19] with active length $l_{EO} = 2$ cm, $V_\pi(0) = 5.1$ V, modulator impedance $Z_m = 47$ $\Omega$, skin effect loss $\alpha_0 = 1.02 \times 10^{-4}$ m$^{-1} \cdot$ Hz$^{-1/2}$, and microwave and optical refractive indices $n_{RF} = 2.15$ and $n_{opt} = 2.14$ with optical loss ignored. The output of the modulator feeds an EO booster with $\alpha = 1$ dB/km, $\beta_2 = -14$ ps$^2$/km, $\gamma = 11.7$ W$^{-1}$km$^{-1}$, attainable in commercial highly nonlinear fibers. The radiation is detected by an ideal photodetector with impedance $Z_d = 50$ $\Omega$ and responsivity $R_d = 1$ A/W. The nonlinear Schrödinger equation is solved by the split-step Fourier method [1], demonstrated to accurately capture nonlinear optical phenomena [14].

In Fig. 3 we examine the normal and boosted link transfer function, $P_{RF,out}(\omega)/P_{RF,in}(\omega)$, as an RF tone's frequency is varied. The MI gain response can be tailored via the EO booster parameters and laser power, here selected for large $V_\pi$ reduction (cf. Fig. 3a) and bandwidth extension (cf. Fig. 3b). In Fig. 3a, a large negative group velocity dispersion is chosen for high-gain at 100 GHz, resulting in over 1000 times larger RF power than the normal link, vastly increasing sensitivity. Choice of smaller dispersion parameter (cf. Fig. 3b) increases the MI bandwidth, enabling near doubling of bandwidth from 144 GHz to 271 GHz.

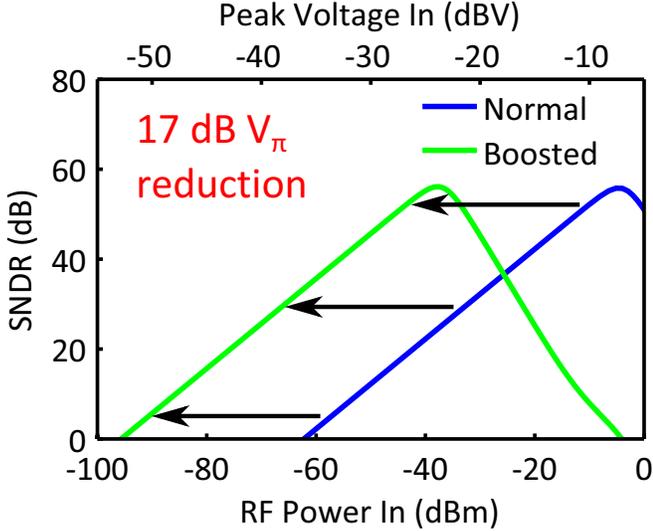

FIG. 4. (color online). Signal to noise and distortion ratio ($SNDR$) (dB) vs. RF Power In (dBm) / Peak Voltage In (dBV) of normal and boosted optical links. Note how the $SNDR$ curve of the boosted link is nearly identical to that of the normal except the input voltage is shifted by 17 dB, and exhibits little to no degradation. This logarithmic shift is consistent with a physical reduction of the characteristic voltage $V_\pi$ by 50 times. Simulation parameters are in the text.

The introduction of an instability into a communication link raises concerns about noise. To be sure, MI has been identified as the source of noise in generation of supercontinuum (white light) radiation [16]. However, such fluctuations arise because in conventional supercontinuum generation, MI is spontaneously triggered by pre-existing noise. Indeed, it has been shown that when MI is stimulated by a deterministic yet minute seed, the output fluctuations are dramatically reduced by as much as 30 dB [14]. In the proposed technique, the modulation sidebands that stimulate MI are phase-locked to the carrier; hence, the process will be inherently stable.

The fidelity of the output RF signal is investigated using the signal to noise and distortion ratio ($SNDR$). Two RF tones, 100 and 101 GHz, are excited in the modulator and the detected photocurrent is bandpass filtered to 95 to 105 GHz with thermal and shot-noise included (cf. Fig. 4). The stability of the MI process stimulated by the deterministic sidebands is reflected in the lack of $SNDR$ degradation for the boosted modulator in Fig. 4. The observed 17 dB shift in the $SNDR$ thus implies over 50 times reduction of $V_\pi$, from 6.7 V to 114 mV.

In summary, we introduced a new concept that exploits modulation instability to enhance the intrinsically weak electrooptic effect, $\chi^{(2)}$, resulting in a highly desired reduction of the characteristic voltage, $V_\pi$, of electrooptic modulators. Simulations demonstrate more than 50 times enhancement of electrooptic effect and modulation sensitivity at frequencies in the 100's of GHz range. Stimulating modulation instability with RF modulation sidebands of an optical carrier enables high frequency low-voltage modulation, an increasingly critical capability that is needed if optical data communication is to keep pace with advances in electronics. Finally, the enhancement in $\chi^{(2)}$ may also find use in wave-mixing applications as long as the frequency translation falls within the MI bandwidth, which can be extended to tens of THz by reduction in the dispersion, $\beta_2$, of the MI medium.

This work was supported by the Office of Naval Research (ONR) and the NSF Center for Integrated Access Networks (CIAN). We thank Robert R. Rice for helpful discussions.

[1] G. P. Agrawal, *Nonlinear Fiber Optics* (Boston, 2007).
[2] T. B. Benjamin, and J. Feir, J. Fluid Mech. **27**, 417 (1967).
[3] H. Elbelrhiti, P. Claudin, and B. Andreotti, Nature (London) **437**, 720 (2005).
[4] R. Bonifacio, C. Pellegrini, and L. Narducci, Opt. Commun. **50**, 373 (1984).
[5] K. Tai, A. Hasegawa, and A. Tomita, Phys. Rev. Lett. **56**, 135 (1986).
[6] K. E. Strecker, G. B. Partridge, A. G. Truscott, and R. G. Hulet, Nature (London) **417**, 150 (2002).
[7] V. Zakharov, and L. Ostrovsky, Phys. D **238**, 540 (2009).
[8] M. Hopkin, Nature (London) **430**, 492 (2004).
[9] C. Kharif, and E. Pelinovsky, Eur. J. Mech. B-Fluid **22**, 603 (2003).
[10] V. E. Zakharov, A. Dyachenko, and A. Prokofiev, Eur. J. Mech. B-Fluid **25**, 677 (2006).
[11] D. Solli, C. Ropers, P. Koonath, and B. Jalali, Nature (London) **450**, 1054 (2007).
[12] D. Solli, C. Ropers, and B. Jalali, Phys. Rev. Lett. **101**, 233902 (2008).
[13] J. M. Dudley, G. Genty, and B. J. Eggleton, Opt. Express **16**, 3644 (2008).
[14] D. Solli, B. Jalali, and C. Ropers, Phys. Rev. Lett. **105**, 233902 (2010).
[15] P. DeVore, D. Solli, C. Ropers, P. Koonath, and B. Jalali, Appl. Phys. Lett. **100**, 101111 (2012).
[16] J. M. Dudley, G. Genty, and S. Coen, Rev Mod Phys **78**, 1135 (2006).
[17] W. S. C. Chang, *RF Photonic Technology in Optical Fiber Links* (Cambridge University Press, 2002).

[18]   A. Chen, and E. Murphy, *Broadband Optical Modulators: Science, Technology, and Applications* (CRC Press, 2011).
[19]   K. Noguchi, O. Mitomi, and H. Miyazawa, J. Lightwave Technol. **16**, 615 (1998).
[20]   V. J. Sorger, N. D. Lanzillotti-Kimura, R.-M. Ma, and X. Zhang, Nanophotonics **1**, 17 (2012).
[21]   S. J. Koester, and M. Li, Appl. Phys. Lett. **100**, 171107 (2012).
[22]   M. Liu, X. Yin, E. Ulin-Avila, B. Geng, T. Zentgraf, L. Ju, F. Wang, and X. Zhang, Nature (London) **474**, 64 (2011).
[23]   D. Solli, G. Herink, B. Jalali, and C. Ropers, Nat. Photonics **6**, 463 (2012).
[24]   M. Droques, B. Barviau, A. Kudlinski, M. Taki, A. Boucon, T. Sylvestre, and A. Mussot, Opt. Lett. **36**, 1359 (2011).
[25]   N. N. Akhmediev, and V. I. Korneev, Theor Math Phys+ **69**, 1089 (1986).
[26]   J. M. Dudley, G. Genty, F. Dias, B. Kibler, and N. Akhmediev, Opt. Express **17**, 21497 (2009).
[27]   M. Taki, A. Mussot, A. Kudlinski, E. Louvergneaux, M. Kolobov, and M. Douay, Phys. Lett. A **374**, 691 (2010).
[28]   G. K. Gopalakrishnan, W. K. Burns, R. W. McElhanon, C. H. Bulmer, and A. S. Greenblatt, J. Lightwave Technol. **12**, 1807 (1994).
[29]   J.-M. Liu, *Photonic Devices* (Cambridge University Press, 2005).
[30]   G. P. Agrawal, *Fiber-Optic Communication Systems* (Wiley, 2012).
[31]   C. E. Shannon, Bell Syst. Tech. J. **27**, 623 (1948).
[32]   P. Kylemark, P. O. Hedekvist, H. Sunnerud, M. Karlsson, and P. A. Andrekson, J. Lightwave Technol. **22**, 409 (2004).